# Adaptive band selection snapshot multispectral imaging in the VIS/NIR domain

Jean Minet[*a], Jean Taboury[a], Michel Péalat[b], Nicolas Roux[b], Jacques Lonnoy[b], Yann Ferrec[c],
[a] Laboratoire Charles Fabry, Institut d'Optique, Univ. Paris-Sud, CNRS, 2 avenue Augustin Fresnel, F-91127 Palaiseau cedex ; [b] Sagem – Groupe Safran – Etablissement de Massy – 178 rue de Paris – 91344 Massy ; [c] ONERA/DOTA – Chemin de la Hunière – 91761 Palaiseau Cedex

## ABSTRACT

Hyperspectral imaging has proven its efficiency for target detection applications but the acquisition mode and the data rate are major issues when dealing with real-time detection applications. It can be useful to use snapshot spectral imagers able to acquire all the spectral channels simultaneously on a single image sensor. Such snapshot spectral imagers suffer from the lack of spectral resolution. It is then mandatory to carefully select the spectral content of the acquired image with respect to the proposed application. We present a novel approach of hyperspectral band selection for target detection which maximizes the contrast between the background and the target by proper optimization of positions and linewidths of a limited number of filters. Based on a set of tunable band-pass filters such as Fabry-Perot filters, the device should be able to adapt itself to the current scene and the target looked for. Simulations based on real hyperspectral images show that such snapshot imagers could compete well against hyperspectral imagers in terms of detection efficiency while allowing snapshot acquisition, and real-time detection.

## 1. THE ISSUE OF SNAPSHOT SPECTRAL IMAGING

Hyperspectral imaging has the potential to detect low contrast targets with minor spectral differences from the background. The central task in hyperspectral system design is to acquire a tridimensional spectral image using a bidimensional image sensor. This is usually done by sequentially acquiring different slices of the spectral image cube. Unfortunately, hyperspectral imagers suffer from two major drawbacks when considering real-time target detection. First, the entire cube is reconstructed from light collected at different times, which raises an issue when dealing with moving scenes, moving platforms or unstable environmental conditions. Second, the high data flow makes real-time data processing difficult. Multispectral imaging solves the second point by reducing the number of bands. However, for military detection applications, it is also mandatory to have a snapshot spectral imager in order to acquire simultaneously the set of images of the same scene with various spectral contents.

Because snapshot spectral imagers must acquire simultaneously all the slices of the spectral cube by parallelizing the measure on one or more image sensors, spectral bands are necessary few yielding to multispectral snapshot imagers. In all snapshot spectral imagers, the spectral separation of light is realized either in the pupil space [1-6] or in the image space [7-12]. Bayer filtering [7] is the simplest method of achieving spectral separation in the image space. Figure 1 gives an example of spectral separation of the light is the pupil space. An array of filters is inserted in the pupil in order to acquire a set of images in a single image sensor, the spectral content of each image being defined by its corresponding filter.

[*] jean.minet@institutoptique.fr ; phone +33 0164533256

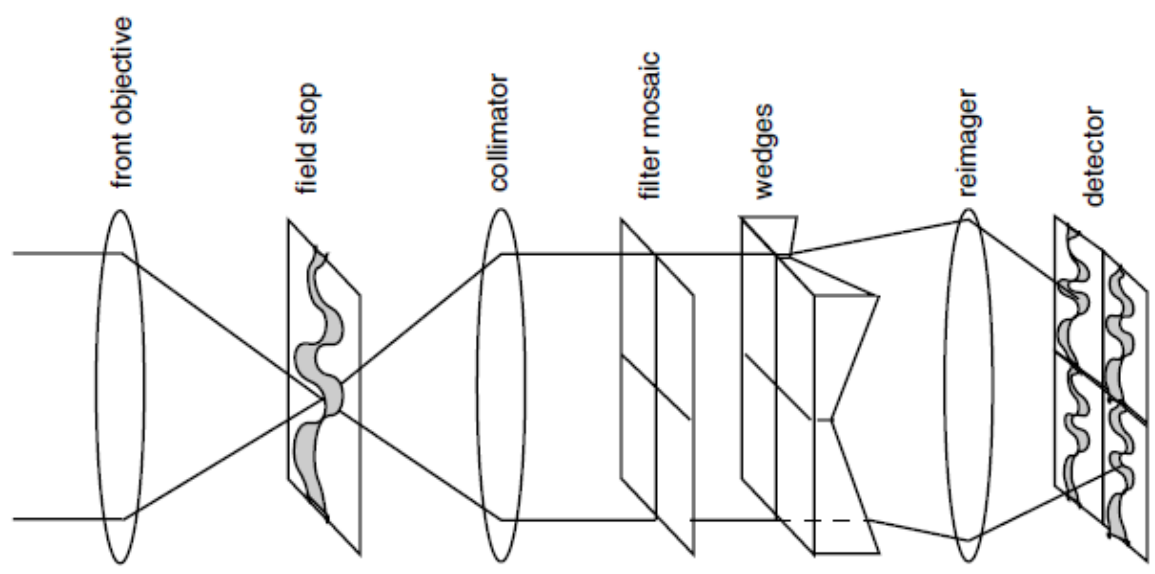

Figure 1. Operating principle of the Simultaneous Multispectral Imager [4]

Whatever the separation method, when designing a snapshot multispectral imager, one has to make a trade-off between the spatial and the spectral resolution i.e. the number of pixels vs. the number of bands. Dealing with high spatial resolution implies acquiring a small number of spectral bands. The spectral profiles of the filters have then to be carefully chosen relatively to the application. This process is usually referred as band selection.

We present a band selection method applied to the detection of spectral targets. This approach can be used to design efficient snapshot spectral imagers based on an array of fixed or tunable filters as the filter mosaic shown in Figure 1. In this paper, we use the method to optimize the spectral profiles of an array of band-pass filters mounted on a snapshot spectral imager. Simulations based on real hyperspectral images show that the proper optimization of the cutoff wavelengths relatively to the current scene and the target looked for can lead to detection efficiencies close to those obtained with hyperspectral resolution.

Section 2 describes our methodology of target detection from hyperspectral images. We use a target spectral library and an atmospheric code compensation to detect the target by means of a matched filter. A criterion of detection efficiency is also defined. Section 3 provides a method to optimize a selection of filters relatively to the criterion of detection efficiency. This is done by the means of a genetic algorithm. Section 4 gives results of band selection in the case of ideal band-pass filters. Section 5 contains concluding remarks and perspectives on future work.

## 2. METHODOLOGY OF TARGET DETECTION

The first step is to define a criterion which measures the performance of the detection of the target when using a given band selection. Algorithms for automatic target detection in hyperspectral imagery have been widely developed in the last decades. Most of these algorithms can be derived from the statistical assumptions which are made on the target and the background surrounding it [13]. We use for simplicity a Matched Filter algorithm which can be proved to be optimal under certain Gaussian assumptions on the target and the background.

The second step is to get hyperspectral images in order to evaluate the detection efficiency of the multispectral imager. We acquire a hyperspectral image of a scene using a *Specim Imspector QE V10E* hyperspectral camera. The scene contains six different paint samples $t_i, i \in \{1,...,6\}$ camouflaged within the scene, each sample being characterized by its true reflectance spectra $\mathbf{s_{ti}}$ measured with a spectroradiometer. The hyperspectral image is made of 256 contiguous spectral bands between 400nm and 1000nm. On each sample, we define a region of interest (ROI) of typically 10 pixels (Figure 2). For each sample $t_i, i \in \{1,...,6\}$; the spectral luminance is integrated on the ROI (Figure 3).

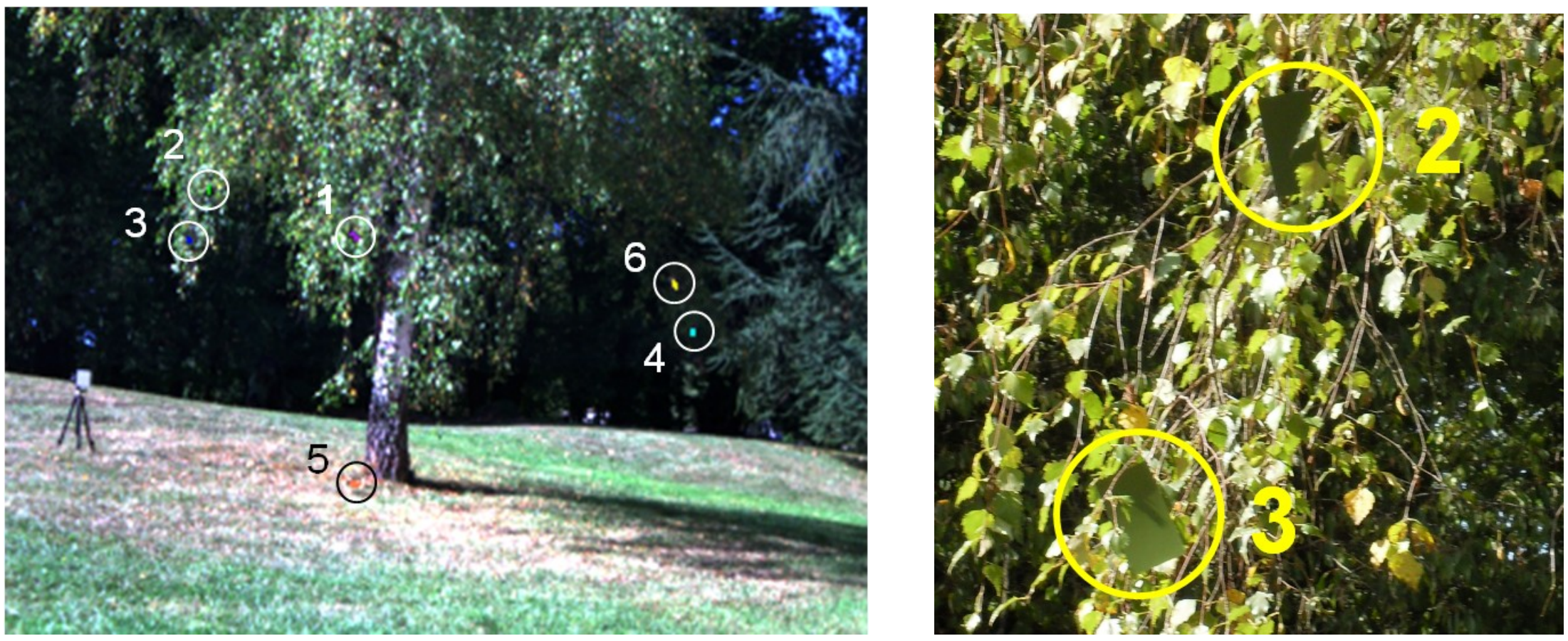

Figure 2. Left: RGB representation of the hyperspectral image of the observed scene. The scene contains six different paint samples for each of which we have defined a region of interest that appears artificially colored on the image. Right: Digital photograph of the paint samples n°1 and 2 as they appear on the scene.

Because each sample has to be detected independently from the illumination conditions, the next step is to get reflectance spectra. A semi-empiric atmospheric code compensation called Quac (quick atmospheric correction) [13] allows us to estimate the reflectance of the scene from the raw hyperspectral image. We note as **x** the reflectance vector of a pixel of the image estimated from Quac. For each sample, we then calculate $\mathbf{m}_{ti} = \langle \mathbf{x} \rangle_{ti}$ where $\langle\ \rangle_{ti}$ represents the mean on the region of interest $t_i$.

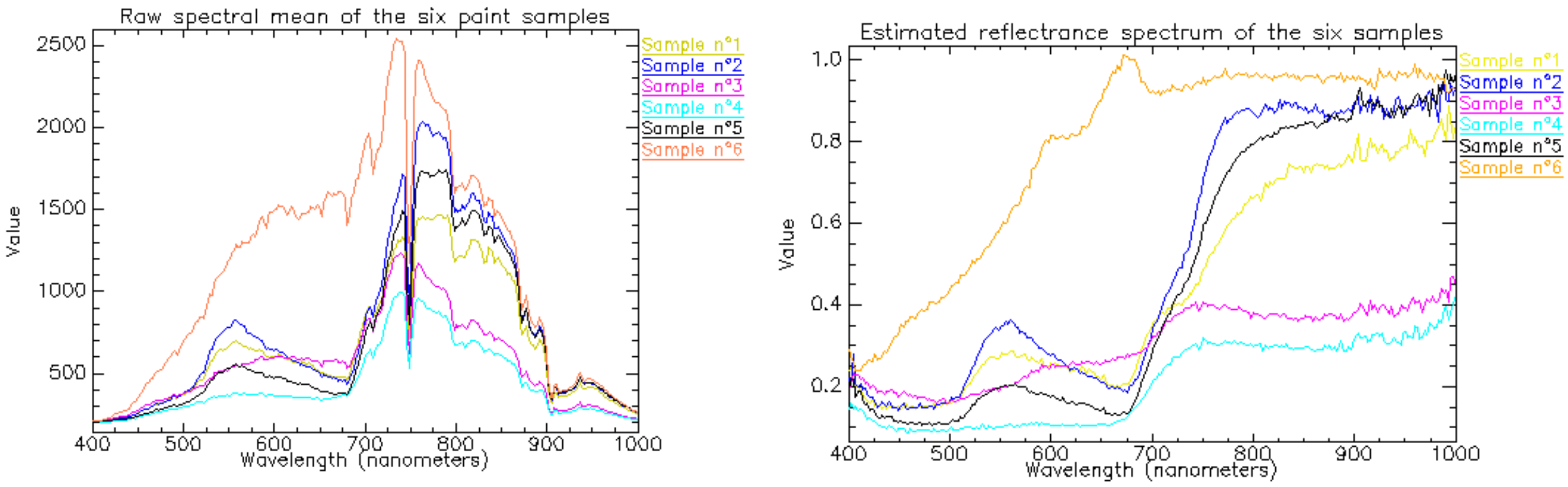

Figure 3. Left: ROI's mean of the six paint samples extracted from the raw hyperspectral image. Right: Estimated reflectance spectrum of the six paint samples from the raw spectral image ( $\mathbf{m}_{ti}, i \in \{1,\ldots,6\}$).

In the same way, we characterize the background by its spectral mean $\mathbf{m}_b = \langle \mathbf{x} \rangle_b$, where $\langle\ \rangle_{ti}$ represents the mean calculated on the region of interest corresponding to the background b; and its covariance matrix $\mathbf{\Gamma}_b = \left\langle (\mathbf{x} - \mathbf{m}_b)^T (\mathbf{x} - \mathbf{m}_b) \right\rangle_b$, where $x^T$ denotes the transpose of the vector x. The sample $t_i$ can be detected using the matched filter:

$$y_{sti} = D_b(\mathbf{s}_{ti}, \mathbf{x}), \tag{1}$$

where $\mathbf{s}_{ti}$ is the true reflectance of sample $t_i$ and where:

$$D_b(\mathbf{u}, \mathbf{v}) = (\mathbf{u} - \mathbf{m}_b)^T \mathbf{\Gamma}_b^{-1} (\mathbf{v} - \mathbf{m}_b), \tag{2}$$

The application of this detector at each pixel location of the estimated reflectance hyperspectral image forms a scalar image called a detection plane. This allows us to define a contrast $C_b(t_j,\mathbf{s}_{ti})$ calculated on the detection plane $Y_{\mathbf{sti}}$ between the ROI of the target $t_j$ and the background b:

$$C_b(t_j,\mathbf{s}_{ti}) = \frac{\left(\langle y_{\mathbf{sti}} \rangle_{tj} - \langle y_{\mathbf{sti}} \rangle_b\right)^2}{\mathrm{Var}\left(y_{\mathbf{sti}}\right)_b}, \tag{3}$$

where $\mathrm{Var}(z)_b$ is the variance of z calculated on the region of interest corresponding to the background b. One can easily show that: $\langle y_{\mathbf{sti}} \rangle_{tj} = D_b(\mathbf{s}_{ti}, \mathbf{m}_{tj})$, $\langle y_{\mathbf{sti}} \rangle_b = 0$ and $\mathrm{Var}\left(y_{\mathbf{sti}}\right)_b = D_b(\mathbf{s}_{ti}, \mathbf{s}_{ti})$. We then derive the contrast:

$$C_b(t_j,\mathbf{s}_{ti}) = \frac{D_b(\mathbf{s}_{ti},\mathbf{m}_{tj})^2}{D_b(\mathbf{s}_{ti},\mathbf{s}_{ti})} = D_b\left(\mathbf{m}_{tj},\mathbf{m}_{tj}\right)\frac{D_b\left(\mathbf{s}_{ti},\mathbf{m}_{tj}\right)^2}{D_b\left(\mathbf{s}_{ti},\mathbf{s}_{ti}\right)D_b\left(\mathbf{m}_{tj},\mathbf{m}_{tj}\right)}. \tag{4}$$

Then using Cauchy-Schwartz inequality, we can show that:

$$C_b(t_j,\mathbf{s}_{ti}) \leq C_b^{max}(t_j) = C_b(t_j,\mathbf{m}_{tj}). \tag{5}$$

This inequality shows that the best contrast is achieved when the measured reflectance spectra $\mathbf{s}_{ti}$ corresponds exactly to the mean spectra $\mathbf{m}_{ti}$ of the target $t_i$ on the estimated reflectance image. Table 2 gives examples of contrasts obtained when detecting the paint samples on the estimated reflectance image. We can note that the target $t_1$ appears with a better contrast on the detection plane $Y_{\mathbf{st3}}$ than on the detection plane $Y_{\mathbf{st1}}$. This is due to the difference between the Quac-estimated reflectance and the true reflectance of the samples.

| $C_b(t_j,\mathbf{s}_{ti})$ | | **ROI on the detection plane** | | | | | |
|---|---|---|---|---|---|---|---|
| | | $t_1$ | $t_2$ | $t_3$ | $t_4$ | $t_5$ | $t_6$ |
| **Detection planes** | $Y_{\mathbf{st1}}$ | **303.7** | 228.3 | 354.7 | 12.2 | 11.8 | 18.1 |
| | $Y_{\mathbf{st2}}$ | 157.3 | **184.3** | 139.4 | 25.2 | 26.4 | 22.3 |
| | $Y_{\mathbf{st3}}$ | 353.9 | 205.2 | **516.1** | 4.9 | 5.6 | 9.7 |
| | $Y_{\mathbf{st4}}$ | 2.1 | 5.1 | 10.9 | **35.2** | 29.6 | 13.6 |
| | $Y_{\mathbf{st5}}$ | 0.4 | 10.5 | 0.1 | 16.4 | **357.8** | 26.1 |
| | $Y_{\mathbf{st6}}$ | 5.5 | 27.4 | 0.1 | 25.5 | 128.9 | **41.3** |

Table 1. Contrasts obtained from the estimated reflectance image when detection is done using true reflectrance spectra $\mathbf{s}_{ti}$. The (i,j) entry contains $C_b(t_j,\mathbf{s}_{ti})$.

| $C_b(t_j,\mathbf{m}_{ti})$ | | **ROI on the detection plane** | | | | | |
|---|---|---|---|---|---|---|---|
| | | $t_1$ | $t_2$ | $t_3$ | $t_4$ | $t_5$ | $t_6$ |
| **Detection planes** | $Y_{\mathbf{mt1}}$ | **420.9** | 185.6 | 603.4 | 2.2 | 1.6 | 6.7 |
| | $Y_{\mathbf{mt2}}$ | 259.0 | **302.2** | 365.3 | 11.0 | 19.0 | 16.1 |
| | $Y_{\mathbf{mt3}}$ | 323.4 | 140.3 | **785.5** | 0.6 | 0.7 | 2.3 |
| | $Y_{\mathbf{mt4}}$ | 11.7 | 42.7 | 5.8 | **74.4** | 59.6 | 23.8 |
| | $Y_{\mathbf{mt5}}$ | 1.4 | 12.0 | 1.1 | 9.7 | **476.1** | 22.7 |
| | $Y_{\mathbf{mt6}}$ | 32.5 | 56.3 | 20.6 | 21.5 | 125.1 | **87.1** |

Table 2. Contrasts obtained from the estimated reflectance image when detection is done using reflectances of the samples in the scene $\mathbf{m_{ti}}$. The (i,j) entry contains is $C_b(t_j, \mathbf{m_{ti}})$.

We can see that the contrast $C_b(t_i, \mathbf{s_{ti}})$ reaches in average 61% of the maximum achievable contrast $C_b^{max}(t_i)$.

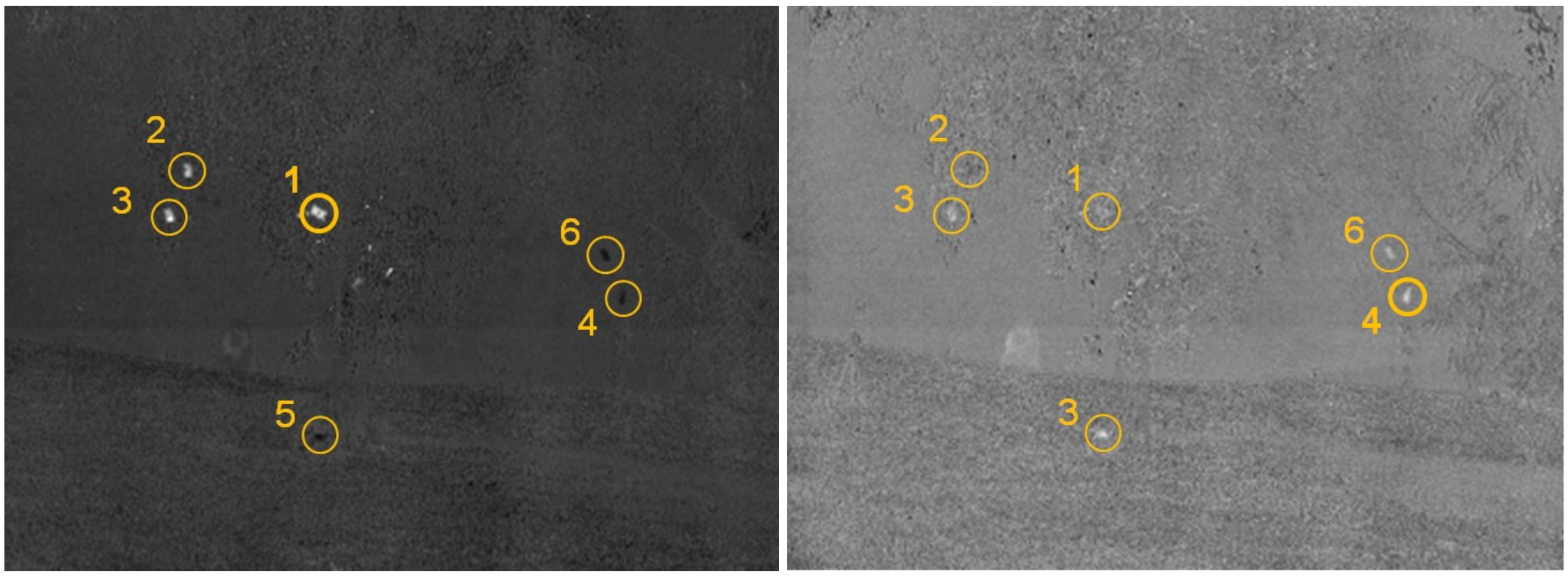

Figure 4. Two detection planes ($Y_{st1}$ and $Y_{st4}$) obtained when detecting two paint samples (left: target n°1, right: target n°4) on the estimated reflectance image from the measured reflectance of the samples to be detected ($\mathbf{s_{t1}}$ and $\mathbf{s_{t4}}$).

## 3. BAND SELECTION METHOD

The process of filtering a spectrum by a set of n spectral filters can be represented by the operation: $\mathbf{y} = \mathbf{R}^T\mathbf{x}$, where $\mathbf{x}$ is the initial spectra (m elements vector), $\mathbf{y}$ is the final spectra (n elements vector) and $\mathbf{R} = (\mathbf{r_1} \ \cdots \ \mathbf{r_n})$ is the $m \times n$ filtering matrix. Its i$^{th}$ column $\mathbf{r_i}$ is an m elements vector representing the spectral profile of the i$^{th}$ filter of the set. The n-bands spectral image can then be used to detect spectral targets. The criterion of detection efficiency (or contrast) becomes:

$$C_{b,\mathbf{R}}(t_i, \mathbf{s_{ti}}) = (\mathbf{s_{ti}} - \mathbf{m_b})^T \mathbf{R}\left(\mathbf{R}^T \mathbf{\Gamma_b} \mathbf{R}\right)^{-1} \mathbf{R}^T (\mathbf{m_{ti}} - \mathbf{m_b}) \tag{6}$$

We define $\widetilde{C}_{b,\mathbf{R}}(t_i, \mathbf{s_{ti}})$ as the normalized version of the contrast:

$$\widetilde{C}_{b,\mathbf{R}}(t_i, \mathbf{s_{ti}}) = \frac{C_{b,\mathbf{R}}(t_i, \mathbf{s_{ti}})}{C_b^{max}(t_i)} = \frac{(\mathbf{s_{ti}} - \mathbf{m_b})^T \mathbf{R}\left(\mathbf{R}^T \mathbf{\Gamma_b} \mathbf{R}\right)^{-1} \mathbf{R}^T (\mathbf{m_{ti}} - \mathbf{m_b})}{(\mathbf{m_{ti}} - \mathbf{m_b})^T \mathbf{\Gamma_b}^{-1} (\mathbf{m_{ti}} - \mathbf{m_b})} \tag{7}$$

We restrain the band selection to a set S of p spectral filters. The band selection problem then consists in selecting the best filtering matrix $\mathbf{R_{max}}$ among the set $S_n = \{(\mathbf{r_1} \ \cdots \ \mathbf{r_n}), \mathbf{r_k} \in S\}$ with respect to the contrast criterion $\widetilde{C}_{b,\mathbf{R}}(t_i, \mathbf{s_{ti}})$:

$$\mathbf{R_{max}}(b, t_i, \mathbf{s_{ti}}, S_n) = \underset{\mathbf{R} \in S_n}{\text{argmax}} \quad \widetilde{C}_{b,\mathbf{R}}(t_i, \mathbf{s_{ti}}). \tag{8}$$

The optimization process can be proved to be a NP-hard problem. This means that one has to test each element of the set $S_n$ to find the optimum solution. If we restrict the problem to a set of 100 spectral filters, the rigorous optimization of the best 10 filters requires more than $10^{13}$ calculations of the criterion $\widetilde{C}_{b,\mathbf{R}}(t_i, \mathbf{s_{ti}})$. We then have to find a mean for estimating an approximate optimum within a restricted time. There are at least two ways of approaching this problem. The first way consists in approximating the resolution of the exact problem by heuristic methods as sequential selection algorithms [12]. The second way consists in finding an approximate problem for which we can obtain an exact solution. This can be done by convex optimization techniques.

We propose to solve our problem heuristically by means of a genetic algorithm. This basically works as a random selection method where the best filters are step by step favored in an evolutionary way. The algorithm runs a population of bits sequences from which we can construct corresponding filtering matrices (Figure 5b).

The population evolves through evolutionary cycles (Figure 5a). Each sequence of the population is evaluated by the contrast $\widetilde{C}_{b,\mathbf{R}}(t_i, \mathbf{s}_{ti})$, where **R** is obtained from the sequence as shown in Figure 5b. The k+1 generation is obtained from the k generation with a reproduction process in which sequences of the n generation are favored as parents (i.e. they are more likely to give rise to children). The offspring is obtained from two parents with a two step process (Figure 5c). In the first step called recombination, we first select the bits shared by both parents and then draw randomly the remaining bits among the bits shared by only one parent; finally a mutation process which consists in randomly change a small number of bits to prevent the algorithm from converging to a local optimum. The k+1 generation is finally obtained by combining the best sequences of the offspring and the best sequences of the k generation.

This reproduction process runs until the convergence of a quality criterion or after a maximum number of evolutionary cycles.

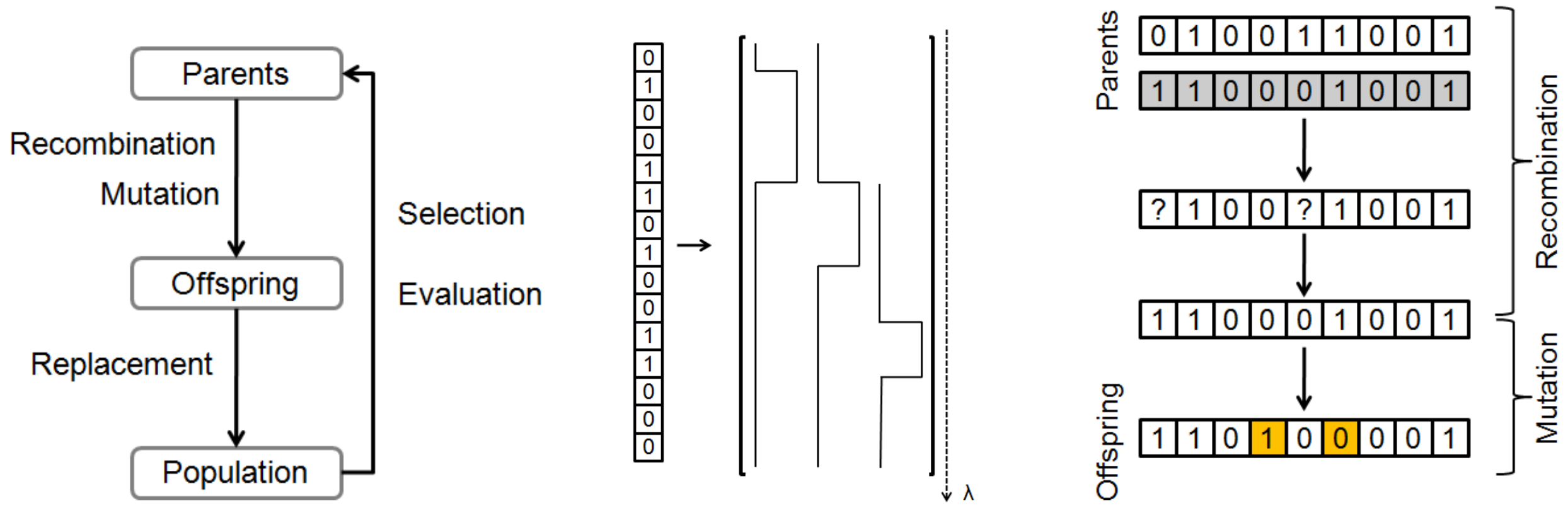

Figure 5. Left (a): Evolutionary cycle. Middle (b): Bit sequence and its corresponding filtering matrix **R** in the case of ideal band-pass filters. Right (c): Reproduction process.

## 4. APPLICATION OF THE BAND SELECTION METHOD WITH IDEAL BAND-PASS FILTERS

We used our band selection method to simulate the detection efficiency of a snapshot multispectral imager based on an array of optimized band-pass filters. We used the hyperspectral image of Figure 2 to simulate the detection efficiency of such an imager. Each band-pass filter is assumed ideal (its response is a rectangular function) and is defined by its two cutoff frequencies which can be tuned as long as there is no overlapping between two filters. There are 256 possible positions for the cutoff frequencies (which correspond to the sampling grid of the hyperspectral image), which means $\binom{256}{2n}$ different filtering matrices can be realized from n spectral filters. For example, there are more than $4.10^{14}$ different combinations of 4 filters.

We segmented the scene of the hyperspectral image into 4 regions corresponding to 4 different types of vegetation in order to simulate 4 different backgrounds $b_i, i \in \{1,\ldots,4\}$. The contrast can then be computed for each of the 6 targets against each of the 4 backgrounds in order to simulate 24 different situations of detection.

For each situation of detection $(i,j) \in \{1,\dots,4\} \times \{1,\dots,6\}$, we optimized the contrast using our band selection process[†]. We obtained the optimized contrast $\widetilde{C}^{max}(b_i, t_j, \mathbf{s}_{tj}, S_4)$:

$$\widetilde{C}^{max}(b_i, t_j, \mathbf{s}_{tj}, S_4) = \max_{\mathbf{R} \in S_4} \widetilde{C}_{bi,\mathbf{R}}(t_j, \mathbf{s}_{tj}). \quad (9)$$

Figure 6 gives the band selection obtained for two such situations. The mean contrast obtained by optimizing independently each of the 24 situations reached 43.6% of the maximum achievable contrast[‡]. We also tried to find the combination which optimizes globally the mean contrast of the 24 situations:

$$\mathbf{R}_{max}^{global}(b, t, \mathbf{s_t}, S_4) = \operatorname*{argmax}_{\mathbf{R} \in S_4} \left\langle \widetilde{C}_{bi,\mathbf{R}}(t_j, \mathbf{s}_{tj}) \right\rangle_{i,j}. \quad (10)$$

We then obtained $\widetilde{C}_{max}^{global}(b, t, \mathbf{s_t}, S_4) = \left\langle \widetilde{C}_{b,\mathbf{R}_{max}^{global}(b,t,\mathbf{s_t})}(t_i, \mathbf{s}_{ti}) \right\rangle_{i,j} = 0.287$. The spectral profiles of the 4 filters of $\mathbf{R}_{max}^{global}(b, t, \mathbf{s_t})$ are given in Figure 7.

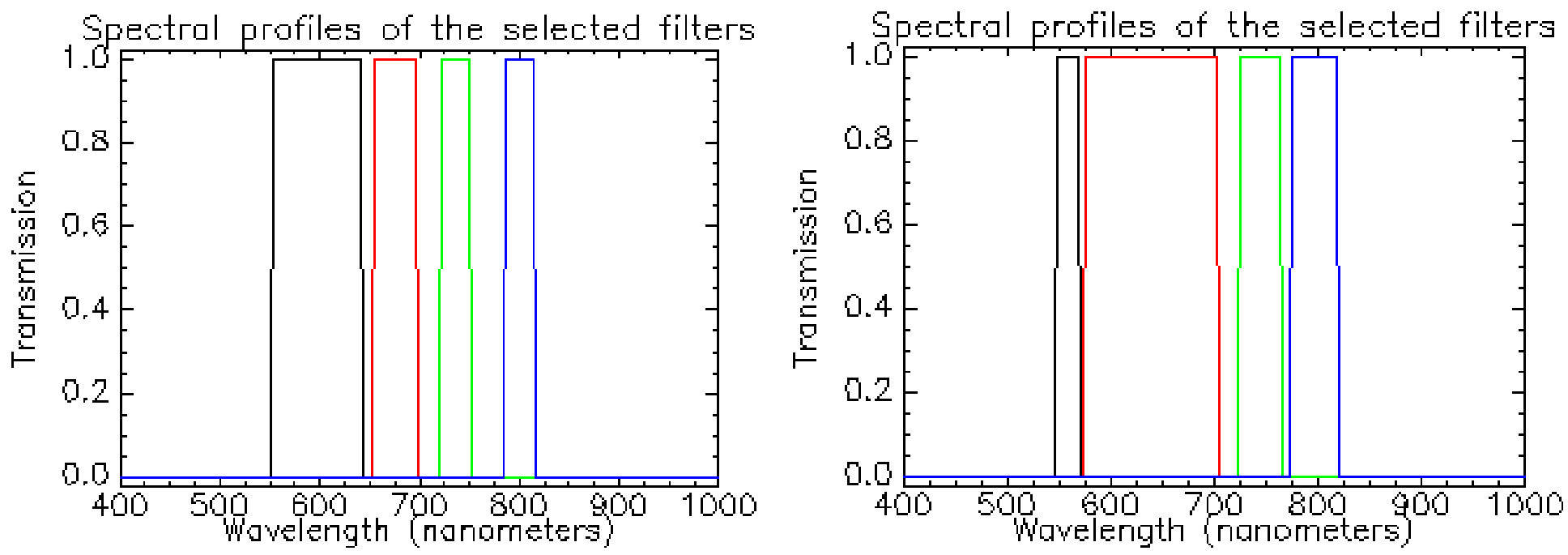

Figure 6. Spectral profiles of the 4 filters of 2 filtering matrices $\mathbf{R}_{\mathbf{max}}(b_1, t_1, \mathbf{s}_{t1}, S_4)$ and $\mathbf{R}_{\mathbf{max}}(b_2, t_1, \mathbf{s}_{t1}, S_4)$ obtained by optimizing the contrast of the target $t_1$ under 2 different backgrounds $b_1$ (left) and $b_2$ (right). The contrast obtained reaches respectively 57.3% and 72.0% of the maximum achievable contrast.

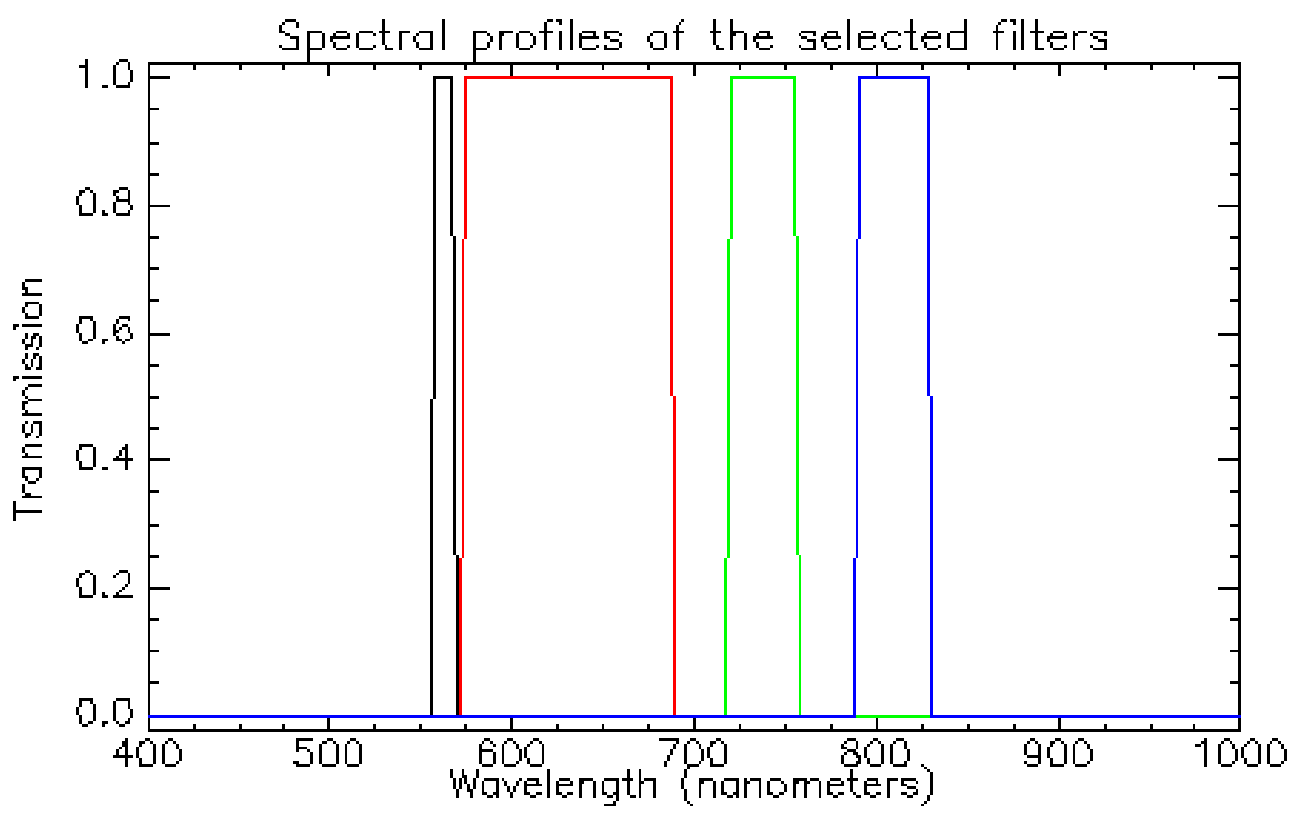

Figure 7. Spectral profiles of the 4 filters of the filtering matrix $\mathbf{R}_{max}^{global}(b, t, \mathbf{s_t}, S_4)$ obtained by globally optimizing the 24 situations (mean detection efficiency of 28.8%).

† The genetic algorithm has been launched with 100 generations of 100 parents.

‡ $\left\langle \widetilde{C}^{max}(b_i, t_j, \mathbf{s}_{tj}, S_4) \right\rangle_{i,j} = 0.436$

Figure 8 represents contrasts when varying the number of bands. The black curve represents the mean contrast obtained when optimizing independently each of the 24 detection situations $\left(\left\langle \tilde{C}^{max}\left(b_i, t_j, \mathbf{s}_{tj}, S_n\right)\right\rangle_{i,j}\right)$. The red curve represents the optimized mean contrast of the 24 situations $\left(\tilde{C}^{global}_{max}\left(b, t, \mathbf{s}_t, S_n\right)\right)$. Then the red curve corresponds to the mean contrast of an imager equipped with fixed filters optimized for the whole context, while the black curve corresponds to the mean contrast of an imager equipped with tunable band-pass filters, which would be able to adapt itself to each detection situation. Figure 8 shows that one can reach 50% of the best contrast achievable with only 5 tunable band-pass filters. In case of non tunable filters, 10 fixed band-pass filters are necessary.

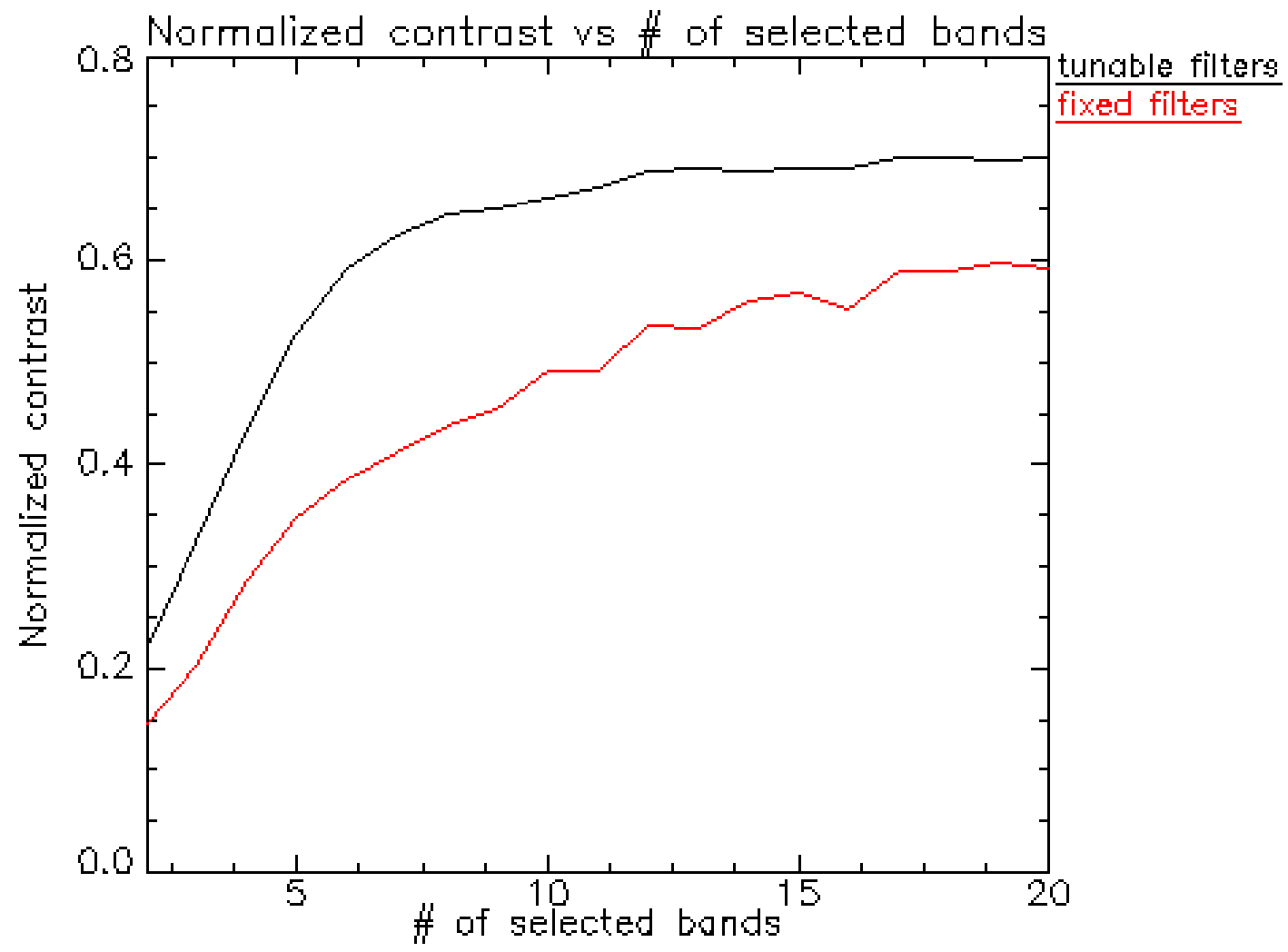

Figure 8. Normalized contrast vs. the number of selected bands.

## 5. CONCLUSION

Hyperspectral imaging becomes a mature technology which has proven its reliability in classification and detection applications. When it comes to real-time applications this technology suffers from its large data rate and it is then often necessary to throw away the major part of the data between the acquisition and the processing. Our band selection approach shows that it is possible to preserve the detecting ability of hyperspectral sensors by acquiring only a small but carefully chosen fraction of the spectral content. It may lead to the design of snapshot multispectral imagers, necessary in case of changing environments. For instance, such imagers can be based on a filter array associated with a lens array. The array of filters can be realized with appropriate interferential filters. The use of tunable filters can be a good way to improve the detection efficiency and the versatility of these snapshot imagers. We are working on the case of Fabry-Perot tunable filters which have already been realized in MEMS technology [16]. The technological realization of such devices looks promising, as well as the theoretical detection efficiency of such devices as shown in Figure 9.

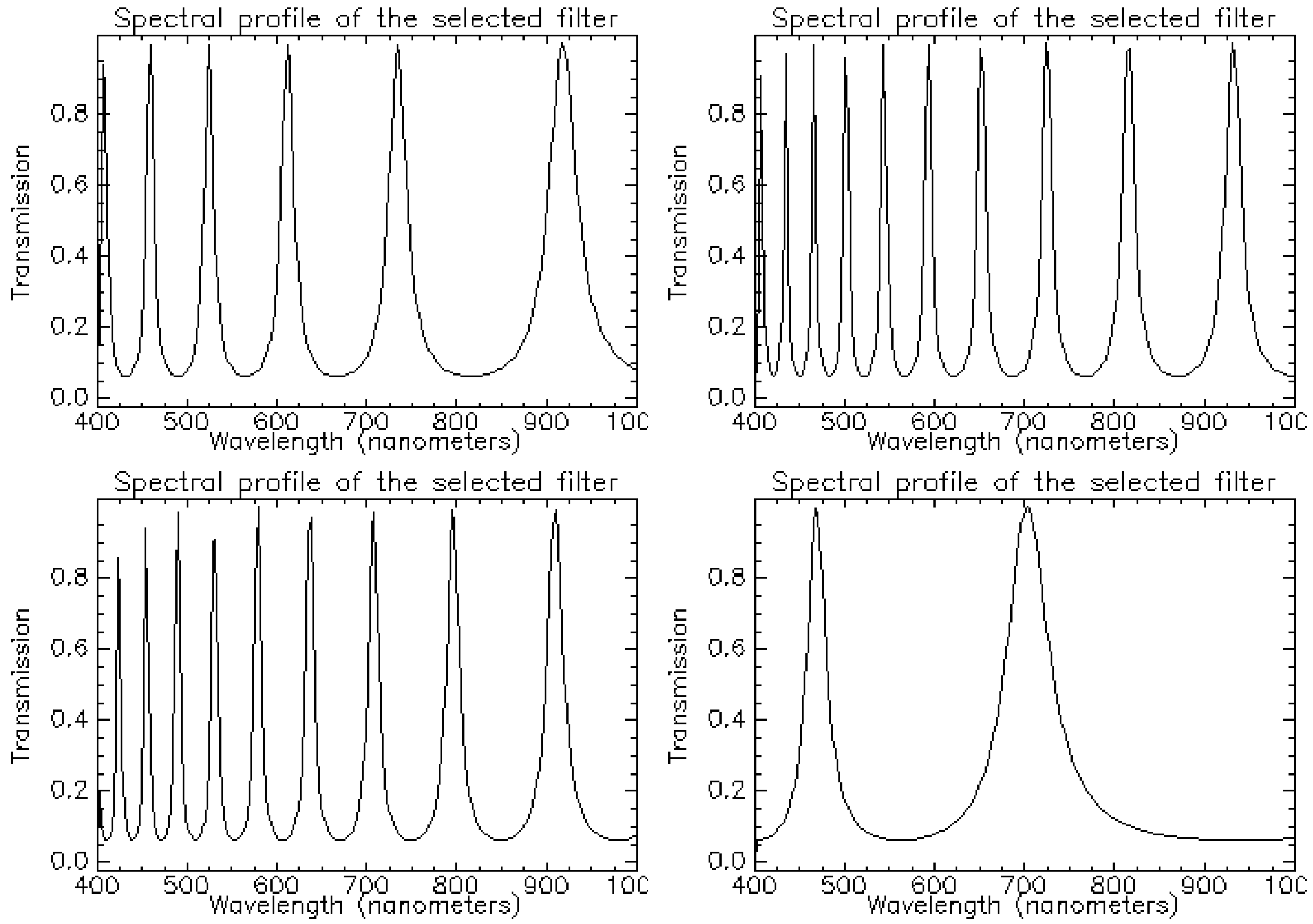

Figure 9. Spectral profiles of the 4 Fabry-Perot etalons of the filter mosaic of Figure 1 obtained by globally optimizing the 24 situations (mean detection efficiency of 26.6%).

## 6. ACKNOLEDGMENTS

We would like to thank Rodolphe Marion of the CEA/DAM/DASE/SLDG/LTSE for the acquisition of hyperspectral images without which we could not have done this work. The work of J. Minet is supported by the Délégation Générale pour l'Armement (DGA/MRIS).